\documentclass[runningheads]{svmult}

\usepackage{makeidx}   
\usepackage{graphicx}  
\usepackage{subeqnar}  
\usepackage{multicol}  
\usepackage{physprbb}  
\usepackage{multind}
\makeindex{s}          
\makeindex{o}          

\begin{document}
\title*{Applications of Indirect Imaging 
\protect\newline
techniques in X-ray binaries}
\toctitle{Applications of Indirect Imaging 
\protect\newline techniques in X-ray binaries }
%
%
\titlerunning{Indirect Imaging techniques in X-ray binaries}
%
\author{Emilios T.Harlaftis\inst{1}}
%
\authorrunning{Emilios T.Harlaftis}
%
%
\institute{Institute of Astronomy and Astrophysics, \\
	National Observatory of Athens,	P.O.Box 20048, \\
	Thession, Athens - 11810, Greece}
\maketitle              

\begin{abstract} 
A  review is given on aspects  of indirect imaging techniques in X-ray
binaries which  are used as   diagnostics tools for  probing the X-ray
dominated accretion  disc physics.  These techniques  utilize observed
properties such as the emission line profile
\index{s}{variability}variability, the  time delays between simultaneous optical/X-ray light
curves, the light curves of eclipsing systems  and the pulsed emission
from  the  compact  object  in order to   reconstruct the  accretion disc's line
emissivity (Doppler tomography),  the \index{s}{irradiation}irradiated
disc and  heated secondary (echo   mapping), the outer  disc structure
(modified eclipse mapping) and the accreting  regions onto the compact
object, respectively.
\end{abstract}

\section{Introduction}

Low-mass  X-ray    binaries  (LMXBs)  involve mass   transfer  from  a
main-sequence  companion  star and accretion  onto a \index{s}{neutron
star}neutron star  or a  \index{s}{black  hole}black hole via  a disc.
Their optical and \index{s}{ultraviolet}UV emission is dominated by
\index{s}{reprocessing}reprocessing          of        \index{s}{X-ray
emission}X-rays,   mainly in  the  disc   but  also on   the companion
star. The fundamental  difference from cataclysmic variable (CV) discs
is the amount of \index{s}{X-ray emission}X-rays produced close to the
accreting region and their \index{s}{irradiation}irradiating effect on
the binary  components.   The X-ray  illumination emanating  from  the
vicinity of  the compact object heats  up the surrounding disc and the
surface of the nearby companion star.  The X-ray heating is so intense
that it controls the radial and  vertical structure of the LMXB discs,
thus overtaking viscous  heating that controls  accretion in CV discs.
The study of LMXBs has been difficult since they  are faint objects in
the optical, and  the  binary  parameters  are  not well   established
because there are very few eclipsing systems and
\index{s}{irradiation}irradiation  is blurring out any  variabilities.
The reader who is interested in more details in
\index{s}{X-ray binary}X-ray binaries can refer to \cite{lewin95} or \cite{wheeler93}. 
Here, we will review a  few examples of  image reconstruction where an
observed property is used  in order to gain  insight into the state of
the companion  star, the accretion disc  and the compact object of the
\index{s}{X-ray binary}X-ray binary.

\section{The heated companion star }

The companion  star in  X-ray binaries is  heated   by the hard  X-ray
radiation, which penetrates the \index{s}{photosphere}photosphere, and
is  re-emitted as blackbody  flux  \cite{hameury86}.  As a consequence
the light re-emitted    by the companion   star depends  on the  X-ray
illumination pattern.  The occultation by the
\index{s}{irradiation}irradiated disc will become apparent as a shadow
on the companion star. The hard X-ray emission that does not encounter
the  accretion  disc and  hits the  star   directly will  heat up  its
\index{s}{photosphere}photosphere  and  cause continuum   emission and
absorption line production.  This orbital heating effect is pronounced
in systems like  \index{o}{Her  X-1}Her  X-1  (where the companion star   is
varying between spectral types $\sim$O9 at maximum to A7/F0 at minimum
\cite{oke76}) and \index{o}{Cyg X-2}Cyg X-2 \cite{casares97})
and can in certain circumstances totally
dominate the star's evolution, as  in e.g.\ the Black Widow pulsar
\index{o}{PSR1957+20}PSR1957+20  \cite{phinney88,ruderman89}. 

That  short period LMXBs might exhibit  such pronounced heating (of an
otherwise cool star) is  supported by the discovery of phase-dependent
HeI absorption  ($\lambda$5876) in  the  secondary  of  the  5.6  hour
eclipsing LMXB \index{o}{X 1822-371}X   1822-371 (see Fig.~1),   which
indicates that  the inner face  of the star  appears to be hotter than
its back face by 10000-15000~K \cite{harlaftis97}.  The heating of the
\index{s}{photosphere}photosphere has never been treated correctly and
any theoretical progress is expected through observational constraints
\cite{dubus99}.  Podsiadlowski \cite{podsliadlowski91} has shown  that
X-ray \index{s}{irradiation}irradiation   can drastically   change the
secondary's structure  (expand atmosphere  by  a factor of   2-3), and
thereby its evolution provided that significant  amounts of energy can
be transported to the  back side \cite{hameury93}. When  applying this
to  \index{s}{outburst}outburst    \index{s}{radial    velocity}radial
velocity data of  the galactic  {\it microquasar}  \index{o}{Nova  Sco
1994}Nova Sco 1994 Shahbaz et  al.  \cite{shahbaz00} found significant
changes to the binary mass solutions.   Orosz \& Bailyn \cite{orosz97}
used a conventional sinusoidal analysis  and derived a black hole mass
of   7.01$\pm$0.22 M$_\odot$ from   a $K$-velocity of 228 km~s$^{-1}$,
whereas Shahbaz  et al.   \cite{shahbaz00}    obtained a much   better
\index{s}{irradiation}irradiated fit to the
\index{s}{radial velocity}radial velocity curve of $K$=215
km~s$^{-1}$ and a   mass  of 5.4$\pm$1.3  M$_\odot$.    The asymmetric
distribution of the absorption line strength  around the inner face of
the companion   star  will  indicate  the  X-ray illumination  pattern
through the disc  and \index{s}{Roche tomography}Roche  tomography may
be used to map the enhanced  absorption line region onto the companion
star  \index{s}{Roche  equipotential}Roche-lobe  surface (Dhillon and
Watson, this Volume).

\begin{figure}[h]
\begin{center}
\includegraphics[width=.5\textwidth]{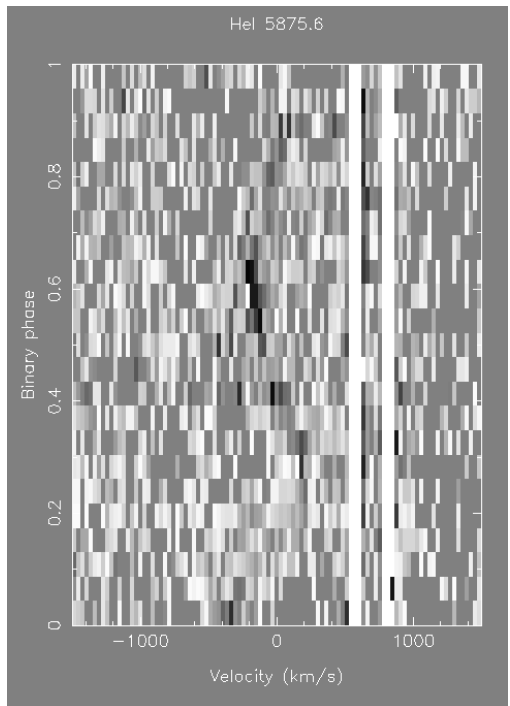}
\end{center}
\caption[]
{The  trailed spectra of  HeI $\lambda$5876 from the binary\index{o}{X
1822-371} X  1822-371.   The  line  is  in  absorption  and  moves from
red-to-blue at phase 0.5  which  is the  signature of the  companion's
star motion.   There   is also an   unidentified  component  at  phase
0.9. Reproduction from \cite{harlaftis97}.}
\label{1822_hei}
\end{figure}

A  new technique which   has started  producing  results,  is ``Echo''
mapping.   This  method utilizes  the \index{s}{time delay}time delays
between optical and X-ray photons in order to map the
\index{s}{irradiation}irradiated   regions   in the   binary   system,
assuming that the  \index{s}{X-ray emission}X-rays are produced at the
centre of  the  disc (O'Brien,  this Volume).    Here,  we present  an
application of the  method on the brightest  X-ray object, the 19-hour
binary system,
\index{o}{Sco X-1}Sco X-1,  which moves along a Z-shaped  curve  in
the     X-ray  colour-colour     diagrams      on  a    $\sim$1    day
\index{s}{timescale}timescale \cite{hertz92}.  It is believed that the
above behaviour reflects changes in the  accretion flow, and therefore
changes in the  structure of the disc (e.g.  thickening of  the disc).
Petro et al. \cite{petro81} found fast ($<$1s) rises in the X-ray flux
followed  immediately by  slower 10-20  second   rises in the  optical
(Fig.~\ref{1822_hei}) at  the flaring branch of  the  Z-curve, a state
which   is  characterized   by    an  enhanced mass    transfer   rate
\cite{hasinger90}.  The flaring branch  is unpredictable and lasts for
only a few hours at most \cite{dieters00}, and many attempts to obtain
simultaneous optical and X-ray  light curves during the flaring branch
have been unsuccessful so far.

\begin{figure}[h]
\begin{center}
\resizebox{\hsize}{!}{\rotatebox{-90}{\includegraphics[width=.6\textwidth]{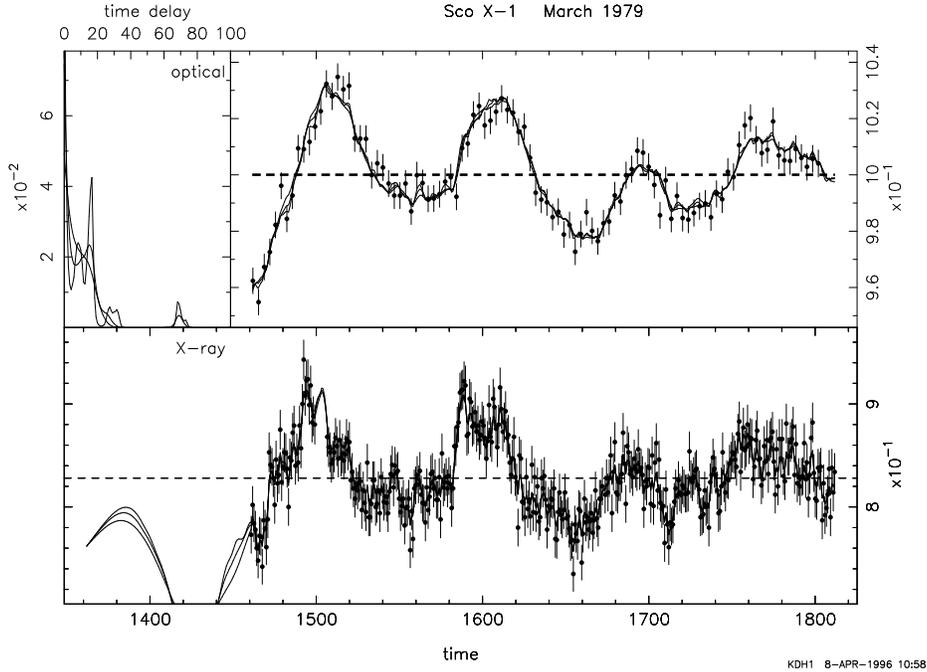}}}
\end{center}
\caption[]{Simultaneous optical  and X-ray light   curves of \index{o}{Sco X-1}
Sco  X-1   \cite{petro81}.  A    re-analysis  of  the  data, using   a
maximum-entropy  technique  \index{s}{entropy}(echo   mapping) gives a
transfer  function (top-left   panel)  which  shows a   \index{s}{time
delay}time delay of $\sim$13 seconds, consistent with
\index{s}{reprocessing}reprocessing off the inner face of the
\index{s}{secondary star}donor star \cite{obrien00}.}

\label{echo1}
\end{figure}

\begin{figure}[h]
\begin{center}
\resizebox{\hsize}{!}{\rotatebox{-90}{\includegraphics[width=.2\textwidth]{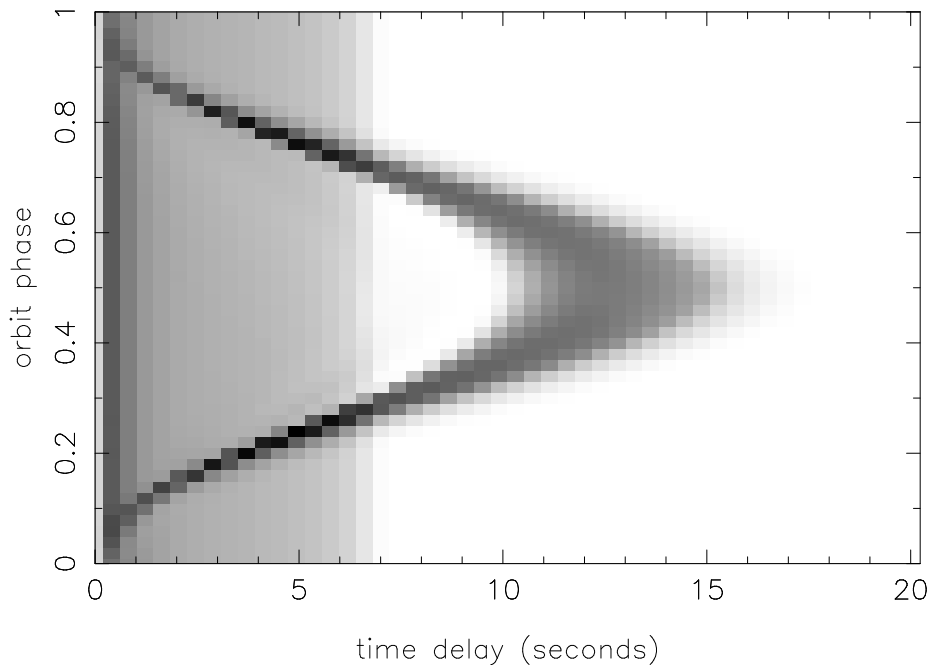}}}
\end{center}
\caption[]
{The echo-phase diagram,   based   on the \index{o}{Sco    X-1}Sco X-1
parameters, which shows   the \index{s}{time delay}time delay   versus
orbital phase.  At orbital   phase  0.5, the range of   \index{s}{time
delay}time delays is maximal (10-15 seconds) \cite{obrien00}.}
\label{echo2}
\end{figure}

\begin{figure}[h]
\begin{center}
\resizebox{\hsize}{!}{\rotatebox{-90}{\includegraphics[width=.05\textwidth]{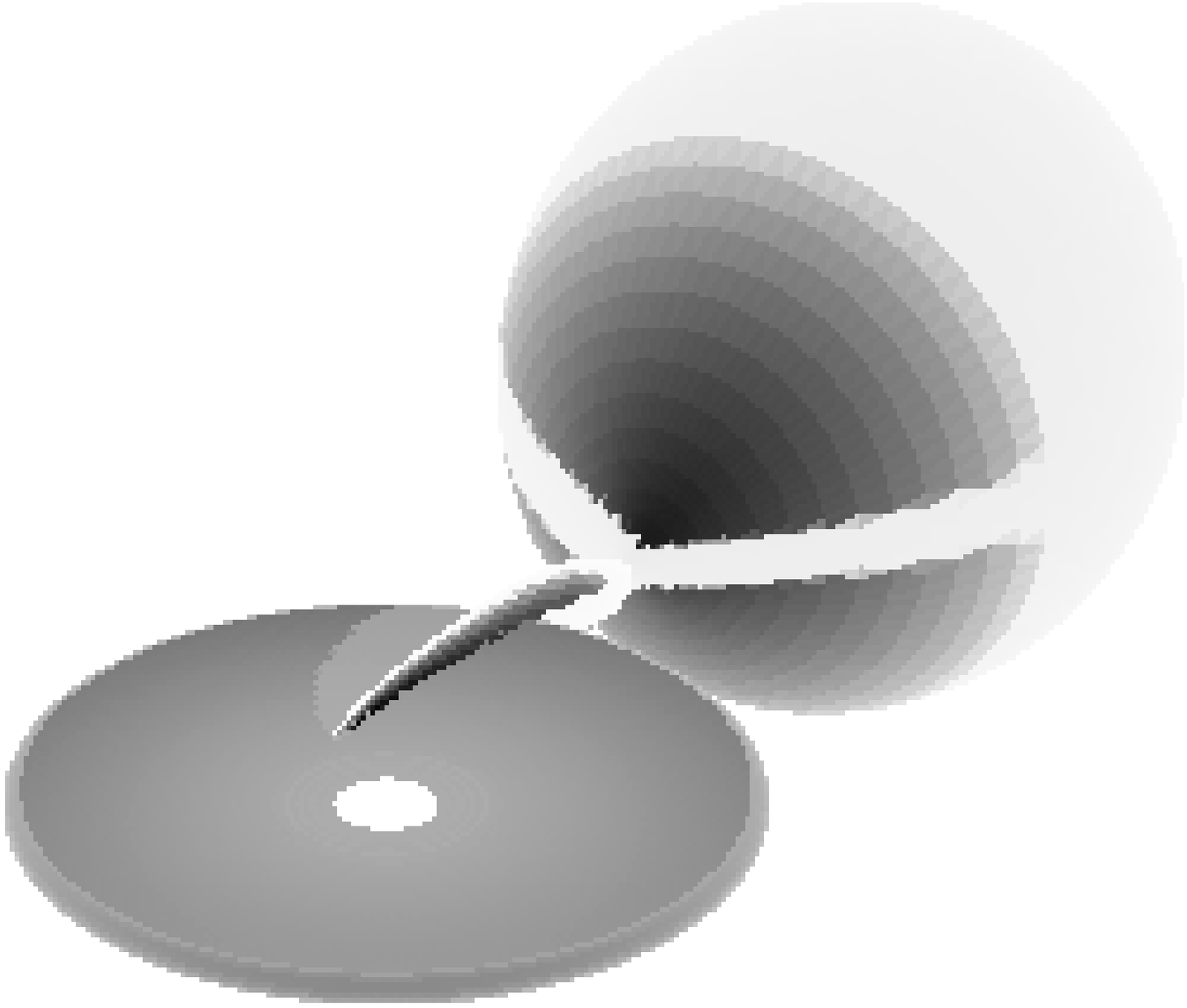}}}
\end{center}
\caption[]
{Model of the \index{s}{irradiation}irradiated system of \index{o}{Sco
X-1}Sco X-1.  The
\index{s}{reprocessing}reprocessing regions \index{s}{irradiation}irradiated by a point-like X-ray source are shown
using the time-delay distribution extracted from the \index{s}{transfer function}transfer function
of the optical and X-ray light curves \cite{obrien00}.}
\label{echo3}
\end{figure}

Therefore,  the only   data   of \index{o}{Sco X-1}Sco  X-1    where a
correlation  between optical and   X-ray photons  is  seen during  the
flaring branch  is  the data  from  \cite{petro81}.   Fig.~\ref{echo1}
shows the simultaneous X-ray and optical light curves of
\index{o}{Sco X-1}Sco~X-1. The transfer   function between the  light
curves (small upper panel in Fig.~\ref{echo1}) shows a distribution of
time-delays which  peak at $\sim$13  seconds  which corresponds to the
light-travel \index{s}{distance}distance of the
\index{s}{neutron star}neutron star to   the inner face   of the
donor  star \cite{obrien00}. A   model diagram showing the  time-delay
versus binary phase   is presented in Fig.~\ref{echo2}.   The constant
time-delay with phase is due to
\index{s}{irradiation}irradiation  in  the disc and   the cut-off at 
6.5 seconds signifies the outer edge of the disc.  The sinusoidal-like
\index{s}{time  delay}time  delays   with   phase  are due    to   the
\index{s}{irradiation}irradiated donor star and is minimum at the back
face but  maximum  at the  inner   face of the   star.   The spread of
time-delays at binary phase  0.5 should eventually constrain the total
area of the heated surface with regard to the
\index{s}{Roche equipotential}Roche lobe.  The model
of the \index{s}{irradiation}irradiated regions in the binary system
of Sco  X-1 is given in Fig.~\ref{echo3}.


\section{ The accretion disc}

\subsection{\index{s}{Doppler tomography}Doppler tomography}

\index{s}{Doppler tomography}Doppler tomography  has  shown  its 
great  diagnostic     value with  the   discovery  of \index{s}{spiral
arms}spiral   shocks    in accretion  discs   (Steeghs,  this Volume).
Application of the technique  in \index{s}{X-ray binary}X-ray binaries
has been  proved  much more difficult,  mainly due  to the   fact that
\index{s}{X-ray binary}X-ray binaries are optically fainter than
\index{s}{cataclysmic variable}cataclysmic variables.  

\begin{figure}[h]
\begin{center}
\resizebox{\hsize}{!}{\rotatebox{-90}{\includegraphics[width=.6\textwidth]{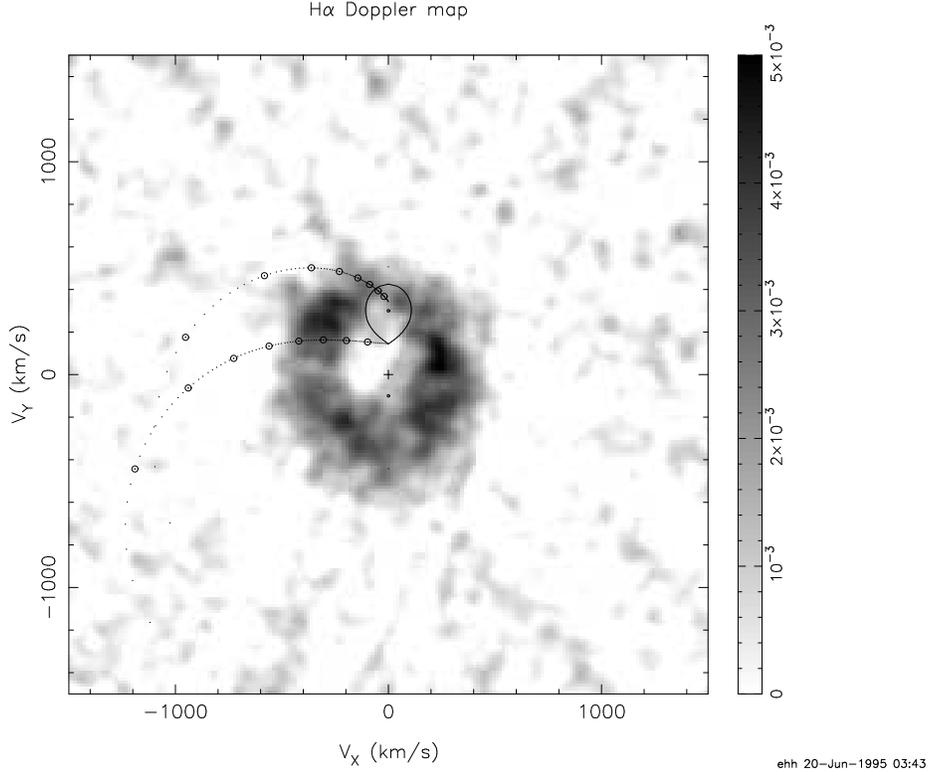}}}
\end{center}
\caption[]
{The H$\alpha$  \index{s}{Doppler map}Doppler map  of X1822-371  shows
the emission-line distribution  as  the  typical  ring-like  structure
\index{o}{X 1822-371} with possibly  enhanced emissivity at phases 0.2
and 0.8.  Reproduction from \cite{harlaftis97}.}
\label{ring}
\end{figure}

For example,   in   persistent X-ray   sources,  such  as  \index{o}{X
1822-371}X~1822-371,  the Balmer  emission  profiles are  not  clearly
double-peaked due to effects which are most likely related to the
\index{s}{irradiation}irradiated,   extended discs  observed  edge-on.
The  trailed  spectra  look  more  like a   blurred version of typical
trailed spectra in
\index{s}{cataclysmic  variable!dwarf nova}dwarf  novae (e.g.\ OY  Car
\cite{harlaftismarsh96}).  The projected   outer rim  of  the disc  is
quite   likely   the    source      of   the   ``blurred''     trailed
spectra. Thereafter, reconstruction of  the emission line distribution
reveals a blurred ring-like   structure with no other clearly  defined
structure. Fig.~\ref{ring} presents such a
\index{s}{Doppler map}Doppler map  of the X-ray persistent source 
 \index{o}{X  1822-371}X~1822-371,  one  of the   {\it  accretion disc
 corona} (ADC)  sources in which \index{s}{X-ray  emission}X-rays from
 the compact object  are not viewed  directly, but are scattered  into
 our line-of-sight by  an extended corona   above the disc  (and which
 explains its  unusually  low $L_X/L_{opt}$ ratio).   Furthermore, the
 discovery of \index{s}{spiral arms}spiral   shocks in the   accretion
 disc of the cataclysmic variable IP Peg\index{o}{IP Peg}
\cite{steeghs97}, which occurred at a time  of a high mass transfer rate
similar to that in LMXBs, raises the question of detecting them in
\index{o}{X 1822-371}X1822-371.  Clearly,  any hint of  disc structure
will    be    easier      to  reveal with       observations   of  the
high-\index{s}{ionisation}ionisation line  He{\small~II} $\lambda$4686
which should provide maps with better clarity than H$\alpha$.  However, He{\small~II}
\index{s}{Doppler map}Doppler maps of other
\index{s}{neutron star}neutron star X-ray  binaries  show complex 
line distributions,  such   as  that  of  \index{o}{XTE  J2123-058}XTE
J2123-058  (Hynes  et al.,  this  Volume; see  also  there  a  list of
\index{s}{Doppler map}Doppler  maps  of \index{s}{neutron star}neutron
star LMXBs).  The latter  \index{s}{Doppler  map}Doppler map shows   a
low-velocity emission at the back-side of the  disc which is difficult
to   interpret.   A  magnetic   \index{s}{magnetic propeller}propeller
scenario is favoured by  Hynes et al.  (this  Volume) as the origin of
the    low-velocity emission.   Alternatively,  it   may be that  this
emission is produced  by the \index{s}{gas stream}gas stream  overflow
crashing back on the  disc, thus the gas  velocities would be  shocked
from around 1200 km s$^{-1}$ to 300 km s$^{-1}$
\cite{harlaftis99a}.

\begin{figure}[h]
\begin{center}
\resizebox{\hsize}{!}{\rotatebox{-90}{\includegraphics[width=.7\textwidth]{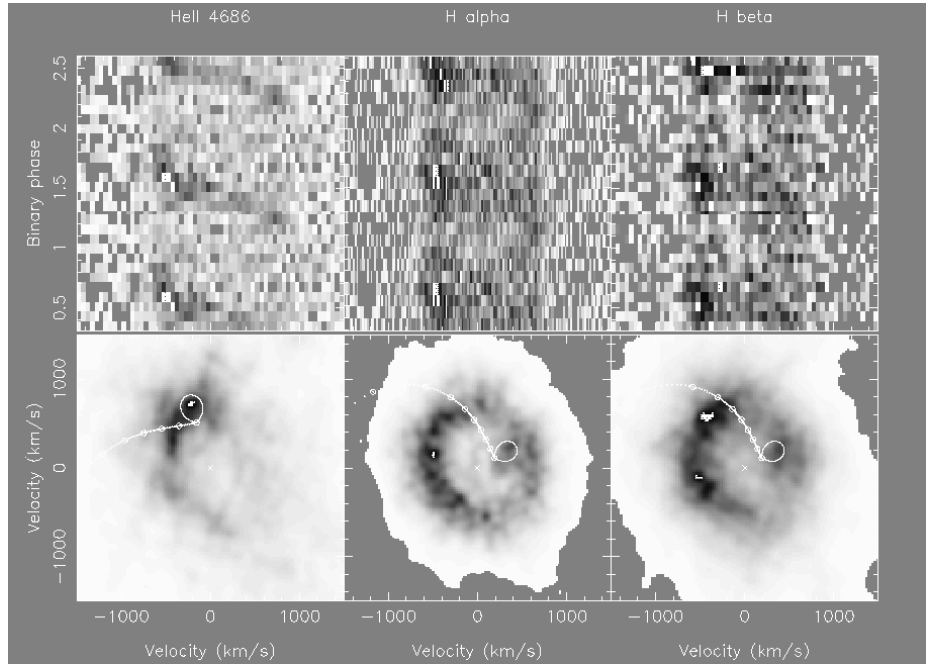}}}
\end{center}
\caption[]
{Doppler images of He{\small~II},  H$\alpha$ and H$\beta$ of the X-ray
transient
\index{o}{GRO J0422+32}GRO J0422+32 during  a mini-\index{s}{outburst}outburst.
Two models  were used to fit  the He{\small~II}  bright-spot with most
likely   the one   that  shows the  gas   stream  passing  through the
bright-spot (the   latter   one  has  a  phase  offset  and  arbitrary
parameters of $K_{c} = 375$ km s$^{-1}$ and $K_{x}  = 75$ km s$^{-1}$;
Harlaftis et  al.  \cite{harlaftis99a}  estimate  $K_{c} =    372$  km
s$^{-1}$    and  $K_{x}   = 43$     km  s$^{-1}$). Reproduction   from
\cite{casares95}.}
\label{j0422}
\end{figure}

\begin{figure}[h]
\begin{center}
\includegraphics[width=.9\textwidth]{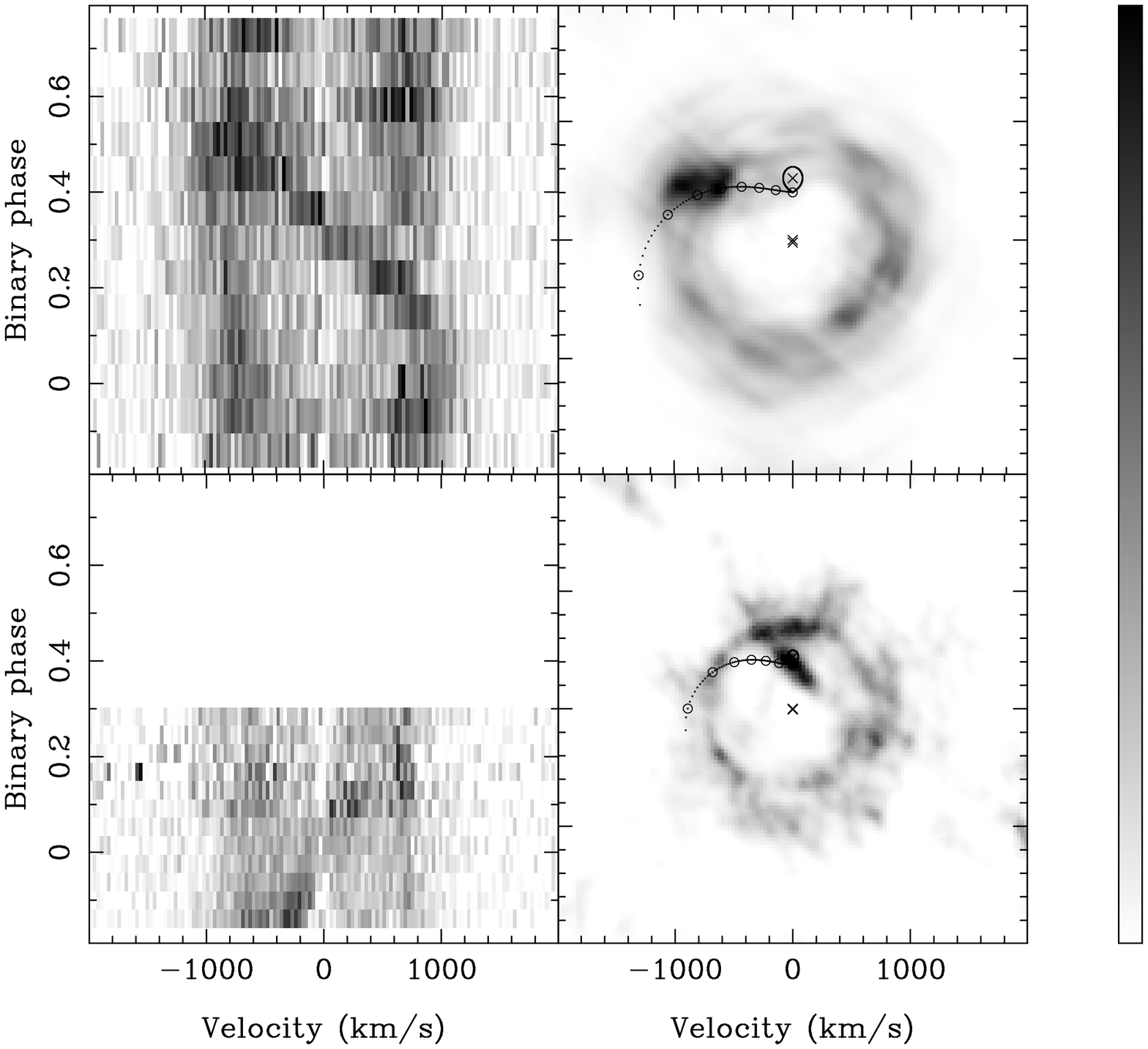}
\end{center}
\caption[]
{The H$\alpha$ Doppler image ({\it  top-right panel}) of the
accretion disc surrounding the  \index{s}{black hole}black hole
GS~2000+25\index{o}{GS 2000+25}  ({\it bottom-right panel} for \index{o}{Nova
Oph 1977}Nova Oph 1977), as reconstructed from 13 Keck-I/LRIS spectra
which are also presented ({\it top-left panel};  {\it bottom-left
panel} for the  12  spectra of \index{o}{Nova Oph 1977}Nova Oph
1977).  By projecting  the image in a particular direction, one 
obtains the H$\alpha$ emission-line profile as  a function of 
velocity; for  example,  projecting toward  the top results in  the
profile at orbital  phase 0.0, which has a blueshifted peak. The path
in velocity coordinates of \index{s}{gas stream}gas streaming from
the dwarf K5 \index{s}{secondary star}secondary star is
illustrated.   The \index{o}{GS 2000+25}GS~2000+25 \index{s}{Doppler
map}Doppler map shows a \index{s}{bright spot}bright spot, at the
upper left quadrant,  which results from collision of the
\index{s}{gas stream}gas stream  with the accretion  disc around the
\index{s}{black hole}black hole. The \index{o}{Nova Oph 1977}Nova Oph
1977 map also shows a trace of an ``S''-wave component which,
however, is not resolved with clarity.  The image was reconstructed
by applying Doppler  tomography,   a maximum
\index{s}{entropy}entropy  technique,   to the phase-resolved
spectra, as described in \cite{harlaftis96,harlaftis97}.}
\label{gs2000}
\end{figure}

In X-ray transient  sources it  has  been difficult to derive  Doppler
maps, since they are very faint at quiescence, and when they are at 
\index{s}{outburst}outburst the binary period is
not accurately known  in order to tailor  phase-resolved observations.
The  disc  outshines the star   and it is  very  difficult to derive a
spectroscopic  period,  and thus a  reliable  ephemeris.  This becomes
apparent with the Doppler \index{s}{Doppler map}tomogram of
\index{o}{GRO  J0422+32}GRO   J0422+32 \cite{casares95}, presented  in
Fig.~\ref{j0422}, where two solutions were  possible at the time given
the uncertainty  in the definition   of absolute phase zero  (inferior
conjunction   of  the companion  star).    However, the  He{\small~II}
emission spot is most likely caused by the impact of the \index{s}{gas
stream}gas  stream   onto the   disc   in   \index{o}{GRO J0422+32}GRO
J0422+32.   When  at   \index{s}{quiescence}quiescence,  the    object
faintness prohibits any phase-resolved  studies.  However, the  advent
of a wealth of X-ray satellites in the 90's  and the advent of the new
generation of telescopes has enabled the first Doppler
\index{s}{Doppler map}tomograms  of the accretion discs around
\index{s}{black hole}black holes. The  line emissivity of such
discs follows a $R^{-b}$ law with $b = 1.5-2.2$
\cite{harlaftis99b}.   The Doppler  map  of \index{o}{GS 2000+25}GS2000+25 in  
\index{s}{quiescence}quiescence  clearly  shows that  there  is
on-going  mass transfer onto the disc  from the presence of the bright
spot   along   the   ballistic trajectory  of      the gas stream   in
Fig.~\ref{gs2000} \cite{harlaftis96}.  However, there is no detectable
emission in the \index{s}{X-ray emission}X-rays or
\index{s}{ultraviolet}UV suggesting that the inner disc may be empty
or frozen (Mukai, private communication).  The same behaviour has also
been observed in \index{o}{A0620-00}A0620-00, i.e.\ a \index{s}{bright
spot}bright  spot in the  outer  disc but no  activity  from the inner
disc~\cite{marsh94,mcclintock995},   a behaviour consistent       with
advection dominated accretion flow models~\cite{eliot99}.  A
\index{s}{Doppler map}Doppler map of  \index{o}{Nova Oph 1977}Nova Oph
1977   in quiescence   -   with  a hint    of some \index{s}{secondary
star}secondary   star     emission   (for    further    details    see
\cite{harlaftissteeghs97}) - is also shown in order to demonstrate the
difficulty   in  building  Doppler    maps  of   X-ray  transients  in
\index{s}{quiescence}quiescence even with 10m class telescopes.

\subsection{The vertical structure of the accretion disc}

\begin{figure}[h]
\begin{center}
\includegraphics[width=.6\textwidth]{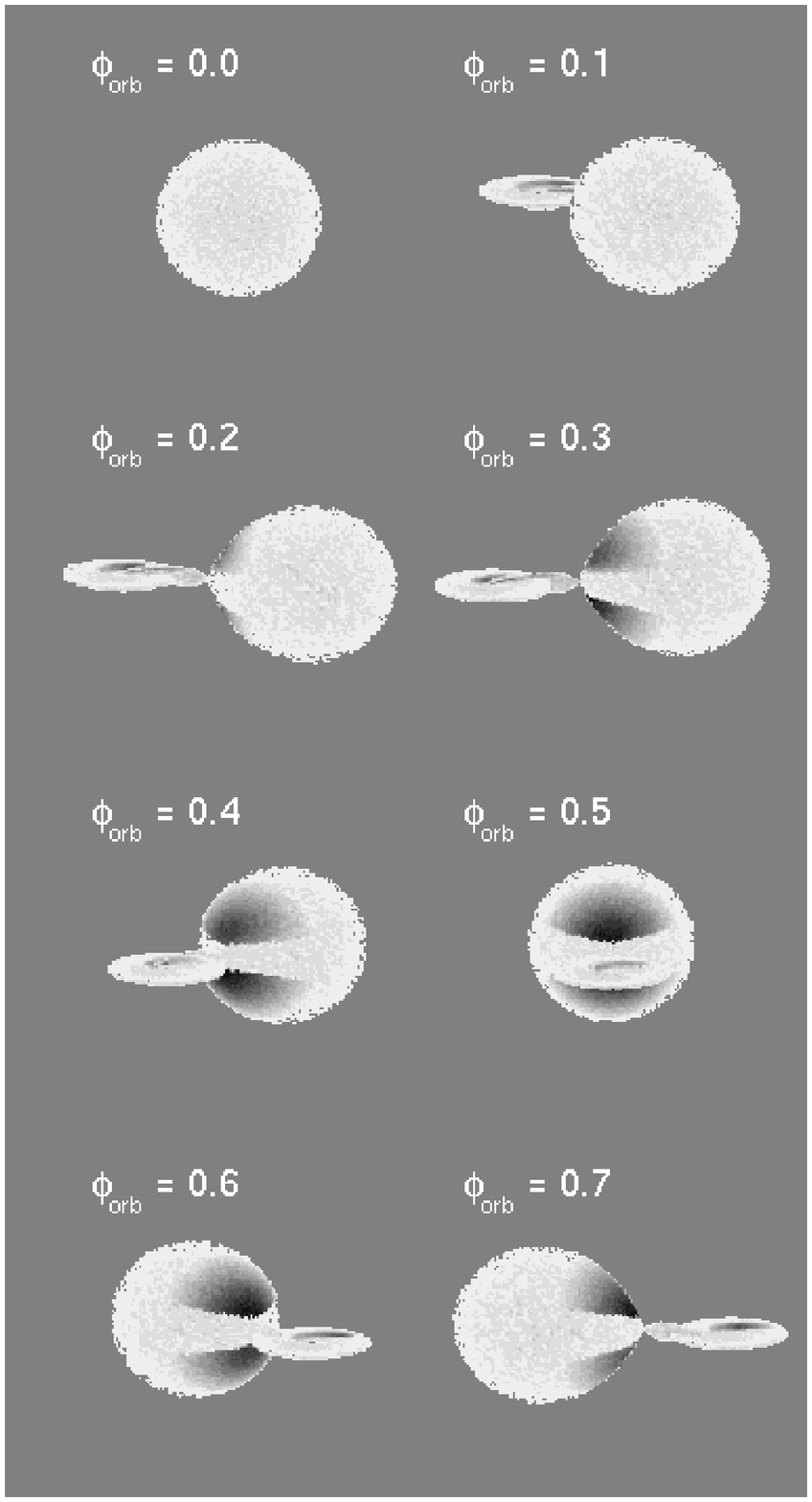}
\end{center}
\caption[]
{Sky projections of the emitting surfaces of \index{o}{Her X-1}Her X-1
over an   orbital cycle. Due    to the large   twist  gradient in  the
accretion   disc,    the  disc  shadow      does  not  display   great
\index{s}{variability}variability over the orbital cycle
\cite{quaintrell98,still97}.}
\label{herx1}
\end{figure}

\begin{figure}[h]
\begin{center}
\includegraphics[width=.8\textwidth]{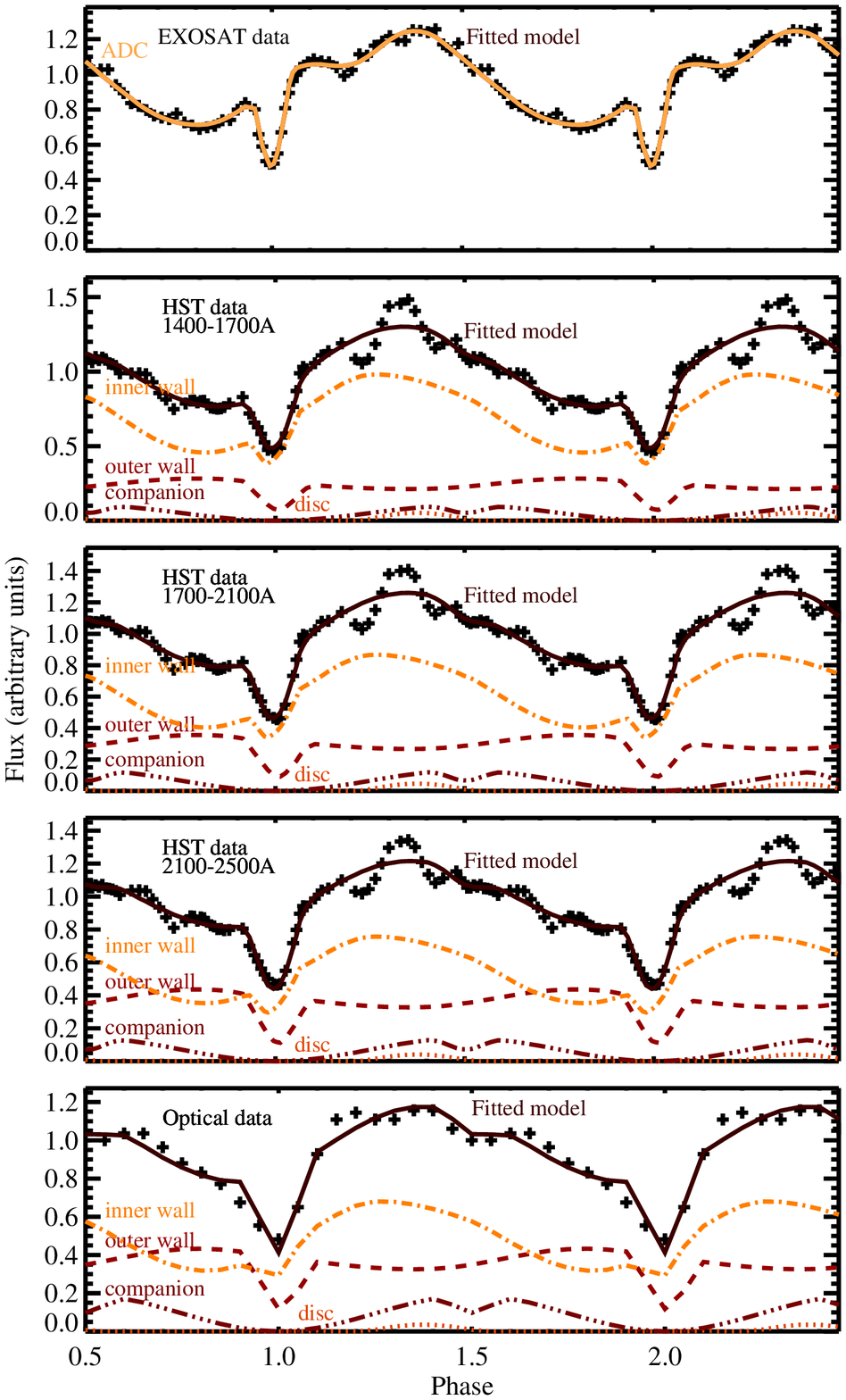}
\end{center}
\caption[]
{The ``dipping''  X-ray source  \index{o}{X 1822-371}X~1822-371 is the
prototype  \index{s}{X-ray  binary}X-ray   binary  for  large vertical
structure, as implicated from the X-ray light curve (X-ray ``dips'' at
binary phase  0.2 and 0.8 and  eclipse).  Simultaneous fits  of a disc
model to  X-ray, \index{s}{ultraviolet}UV  and  optical  light curves.
The \index{s}{ultraviolet}UV-HST data have  been separated into  three
bands.The contribution of each model  component (companion star, disc,
outer and   inner disc   wall)  to each   curve  is  shown  separately
\cite{puchnarewicz95}.}
\label{1822_x}
\end{figure}

Although       many    measurements   of   the     \index{s}{accretion
disc!radius}radial disc structure exist, mainly from \index{s}{eclipse
map}\index{s}{eclipse    mapping}eclipse  mapping    studies  of   the
temperature-radius relationship (e.g.\ \cite{rutten93}), very   little
is known  about the vertical stratification  of  accretion discs.  The
vertical structure of the  disc would require a temperature  inversion
to explain, for example, the emission lines from discs. Indeed, Hubeny
\cite{hubeny94}    finds that the  emergent   spectrum  depends on the
vertical  structure model and  constraints on  the  value of  the disc
\index{s}{viscosity}viscosity could be   imposed from measurements  of
the optical thickness of the disc lines.  In
\index{s}{irradiation}irradiated discs,  the vertical height goes like  $H/R  \sim
(R^{-1/8}-R^{-2/7})$  which results in   a concave disc at large
\index{s}{distance}distances from the compact  object rather than a
dependence  like $H/R   \sim  R^{-1/8}$ in  a  viscously-heated  disc.
Analysis of existing X-ray and
\index{s}{ultraviolet}UV light curves of \index{o}{X
1822-371}X~1822-371 (Fig.~\ref{1822_x}) requires a $H/R$ ratio for the
disc that is rather large ($\sim$0.2;
\cite{puchnarewicz95}). The
question  that arises is  how one  can utilize the vertically-extended
accretion  discs in order to  derive constraints of the accretion disc
properties.  One (new) way is   to  observe simultaneous optical   and
X-ray light curves and analyse them using
\index{s}{echo mapping}echo mapping.  The analysis should reveal
\index{s}{X-ray reprocessing}X-ray reprocessing  regions in the disc, as  in the
case of \index{o}{GRO J1655-40}GRO J1655-40 (\cite{hyneshaswell98} and
O'Brien in this Volume). In principle, these regions should eventually
constrain  the vertical structure with azimuth.   The other  way is to
infer the outer \index{s}{accretion disc!thickness}disc thickness from
the X-ray shadow cast on the companion star.  The 1.24 seconds pulsar
\index{o}{Her X-1}Her X-1 \index{s}{irradiation}irradiates the
accretion disc  around  the  \index{s}{neutron  star}neutron  star and
gives rise to a precessing  and warped accretion  disc.  The shadow of
the accretion disc onto the inner face of  the companion star provides
a  diagnostic   of  the  vertical    structure   of the   disc    (see
Fig.~\ref{herx1}; also see \cite{harlaftis99c}).

\begin{figure}[h]
\begin{center}
\resizebox{\hsize}{!}{\rotatebox{-90}{\includegraphics[width=.7\textwidth]{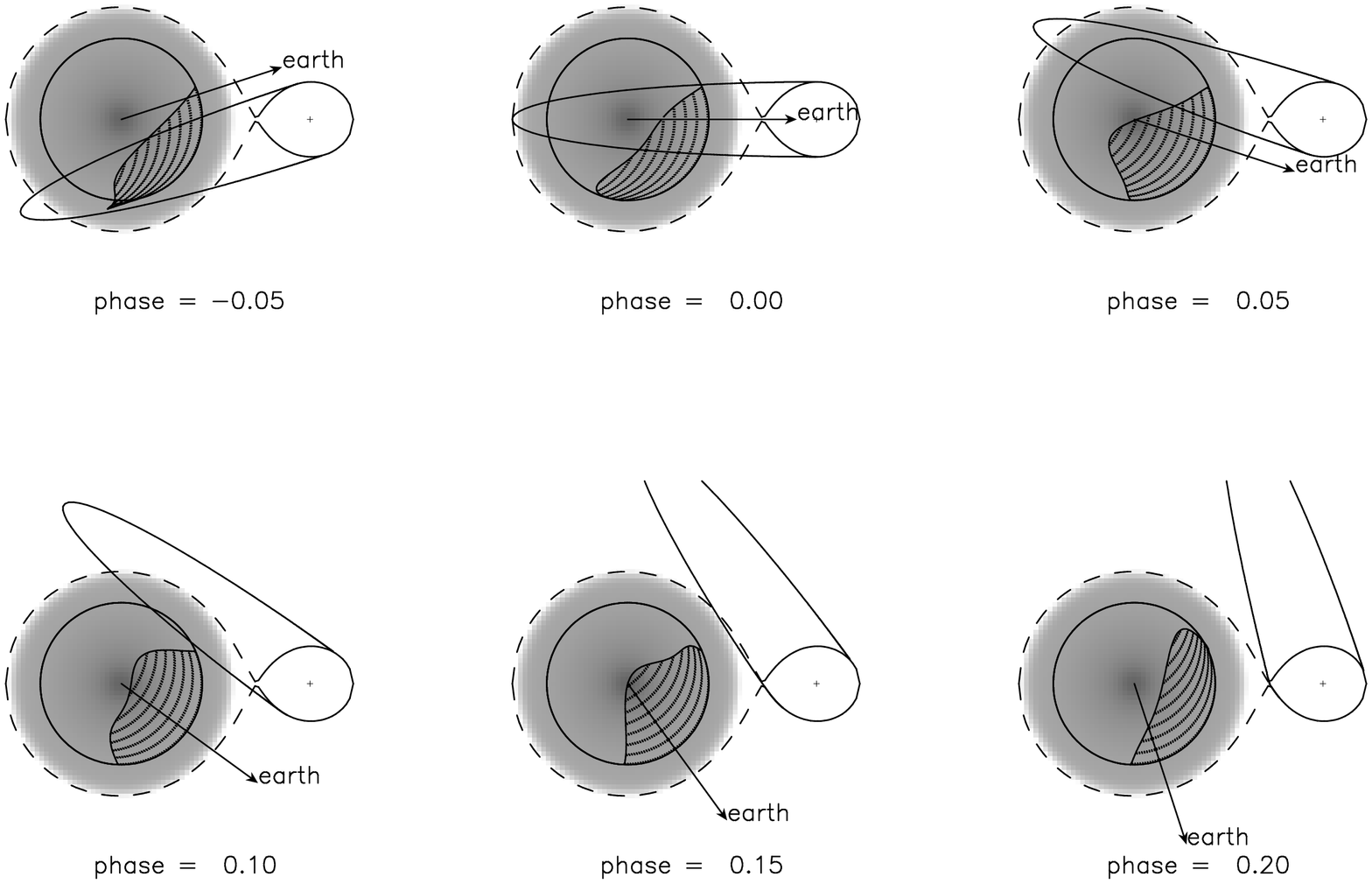}}}
\end{center}
\caption[]
{The  model  for  fitting  the  \index{s}{ultraviolet}UV  dips in  the
super\index{s}{outburst}outburst light  curves  of \index{o}{OY Car}OY
Car is triggering structure in the outer  disc which evolves with time
on a  dynamical  \index{s}{timescale}timescale. The areas of  the disc
surface obscured by the rim and the \index{s}{secondary star}secondary
star are illustrated.   The rim rotates in  the binary frame  and each
rim  element starts to  \index{s}{flare}flare up  at the same position
relative to the \index{s}{secondary star}secondary star. The centre of
the disc is eclipsed by the \index{s}{secondary star}secondary star at
phase  0.0  and by the  rim  between phase  0.05 and  0.15 causing the
\index{s}{ultraviolet}UV dip
\cite{billington96}.}
\label{uvdip}
\end{figure}

Another    way  is  to   extend  the    traditional  \index{s}{eclipse
map}\index{s}{eclipse mapping}eclipse mapping technique (see Baptista,
this Volume) in  the steps of Rutten  \cite{rutten98} in order  to fit
the full orbital light curve of  a prototype vertically-extended disc.
Indeed, Billington et al.
\cite{billington96} explained the  \index{s}{ultraviolet}UV dips
seen   in   the  light  curves  of     \index{o}{OY Car}OY Car  during
super\index{s}{outburst}outburst as outer rim structure where
\index{s}{ultraviolet}UV light is reprocessed  into optical, using 
the model shown  in Fig.~\ref{uvdip}.   This modified eclipse  mapping
technique fitted  the light  curves  by  varying  both  the disc  flux
distribution and the  outer  rim  structure.  The reconstructed    rim
structure  dependence with binary phase  is presented in the following
figure (Fig.~\ref{rim1}). The rim  arises at  the  outer disc,  and in
particular from radii larger than 0.55 $R_{L_{1}}$.  This is reflected
in Fig.~\ref{rim2} where  the rim at 0.5  $R_{L_{1}}$ has a  different
structure than the other outer rims. This is because the rim structure
is too far inside the disc and this forces  the re-distribution of the
flux to an   artificially asymmetric flux  map.   A map which is   not
consistent     with  the   observed   axisymmetric    discs   in   the
\index{s}{ultraviolet}UV.   The  variation of  the mean  rim height with
azimuth in
Fig.~\ref{rim1} matches  the wavelength dependence  of the  absorption
coefficient  of  a hot  disc atmosphere  of 10,000  K  (except for the
shorter wavelengths  where the line emission  is not dominated anymore
by the disc but by a  wind).  The success of  this model in explaining
the \index{s}{ultraviolet}UV dips which appear simultaneous to the 
\index{s}{superhump}superhump maximum during
super\index{s}{outburst}outburst of OY Car (outer disc structure where
\index{s}{ultraviolet}UV light is reprocessed into the optical) points
now  to a  revision of the  technique  and application to the  complex
light  curves   of   the   prototype  of   vertically-extended  discs,
\index{o}{X 1822-371}X 1822-371, as the next step.

\begin{figure}[h]
\begin{center}
\includegraphics[width=.6\textwidth]{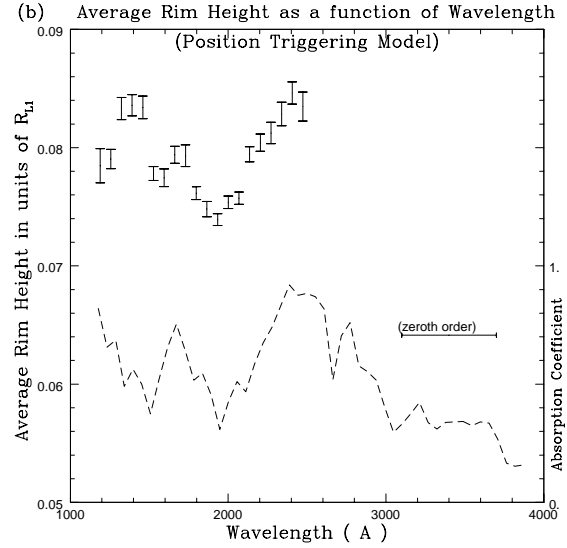}
\end{center}
\caption[]
{The average rim height as a  function of wavelength. The lower dashed
line  and the  right-hand  scale show the   theoretical  opacity of an
accretion disc atmosphere at 10,000 K for the same wavelength band.}
\label{rim1}
\end{figure}

\begin{figure}[h]
\begin{center}
\includegraphics[width=.6\textwidth]{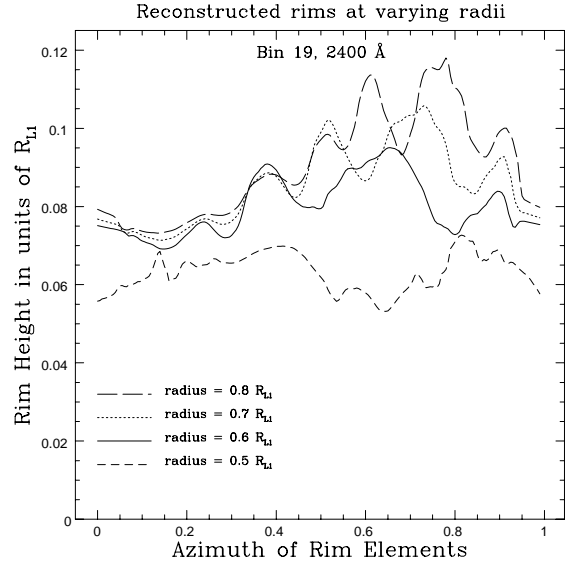}
\end{center}
\caption[]
{The reconstructed rims at radii  0.6 R$_{L_{1}}$, 0.7 R$_{L_{1}}$ and
0.8 R$_{L_{1}}$. The rim at 0.5 R$_{L_{1}}$  is not well reconstructed
as is evidenced by the different shape of  the rim at this radius. The
discrepancy on  the left side of the  curves  is due to  non-disc line
emission.}
\label{rim2}
\end{figure}

\clearpage

\section{\bf The compact object}

\begin{figure}[h]
\begin{center}
\includegraphics[width=1.1\textwidth]{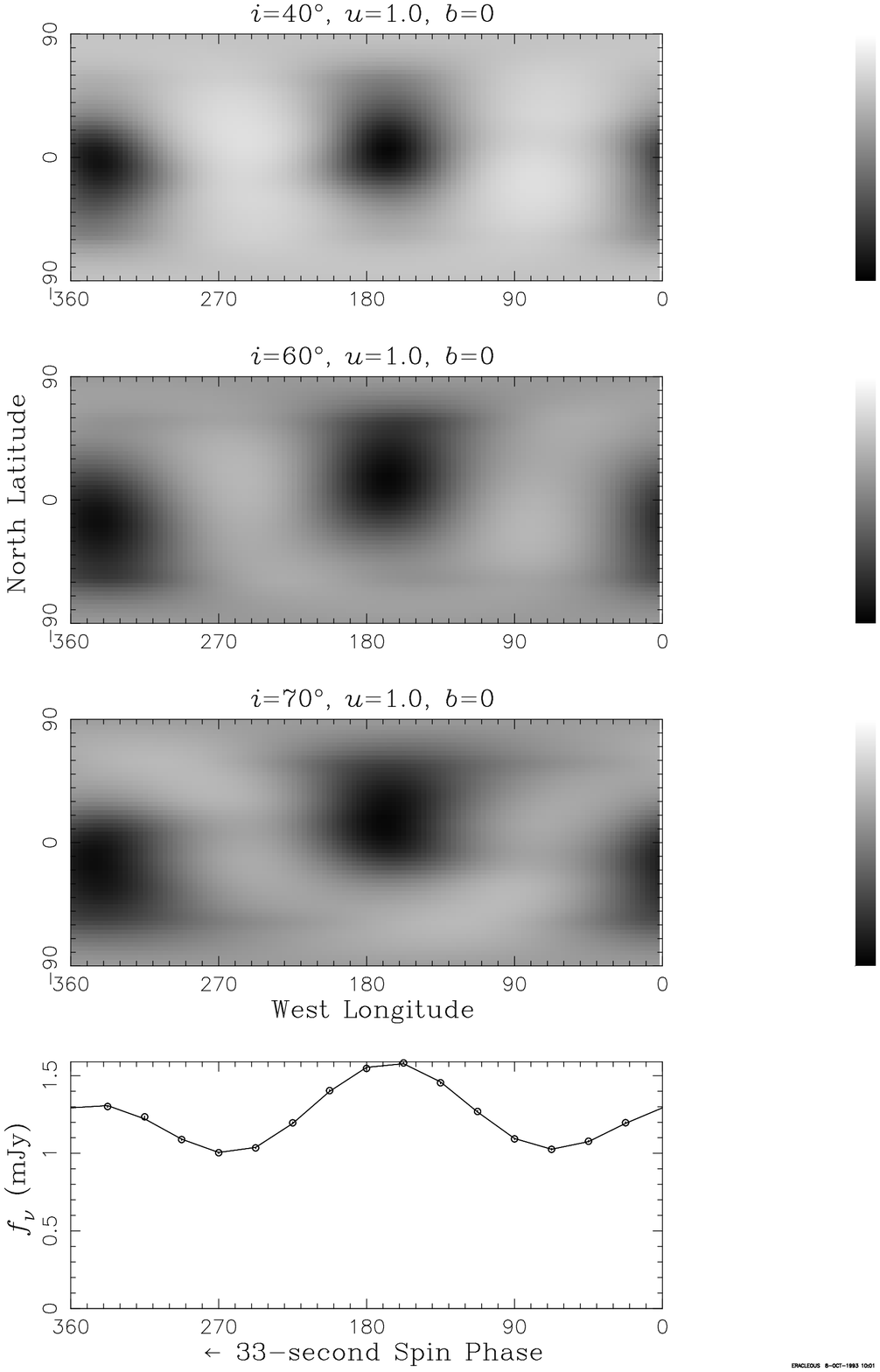}
\end{center}
\caption[]
{Maximum    \index{s}{entropy}entropy  maps   of  the  \index{s}{white
dwarf}white dwarf surface  brightness distribution reconstructed  from
the observed   \index{s}{ultraviolet}UV  pulse  profile   showing  the
emission spots on the  surface of the rapidly rotating \index{s}{white
dwarf}white dwarf at an
\index{s}{inclination}inclination of $60^\circ$, \index{s}{limb
darkening}limb darkening  coefficient  of one  and  background light
of zero\index{o}{AE Aqr} \cite{eracleous94}.}
\label{tomo1}
\end{figure}

The compact object hidden in the luminous \index{s}{X-ray binary}X-ray
binary can  be inferred  by using various  techniques  (except for the
classical radial velocity study of the wings of the emission lines
\cite{filippenko95}).
The most  interesting are based  on the observed coherent pulses which
must arise either from the  surface of the compact  object or from the
coupling region between the accretion disc and the
\index{s}{magnetosphere}magnetosphere.             Indeed,         the
\index{s}{ultraviolet}UV  continuum double  pulse profile observed  at
the  33-sec spinning period of the   compact object in the cataclysmic
variable \index{o}{AE   Aqr}AE  Aqr was successfully    modeled as the
accreting spots on the surface of the compact object (Fig.~\ref{tomo1}
\cite{eracleous94};  but see also work  by \cite{cropper94} who mapped
the accreting regions onto the magnetic white dwarf of  ST LMi).  This
is  also  the    system  where    the   magnetic    \index{s}{magnetic
propeller}propeller model  has found substantial support from emission
line observations (see Wynn, this Volume).  According to the magnetic
\index{s}{magnetic   propeller}propeller   model,   the compact object
rotates  so fast that   the gas  cannot accrete  on  it but  rather is
propelled away.  This concept, first proposed for neutron star
\index{s}{X-ray binary}X-ray binaries in the 70's 
\cite{illarionov75}, has  recently returned  as a  potential 
model for  \index{s}{neutron star}neutron star X-ray binaries (Hynes
et al., this Volume), after it found sound support
in the AE Aqr case.

\begin{figure}[h]
\begin{center}
\includegraphics[width=.6\textwidth]{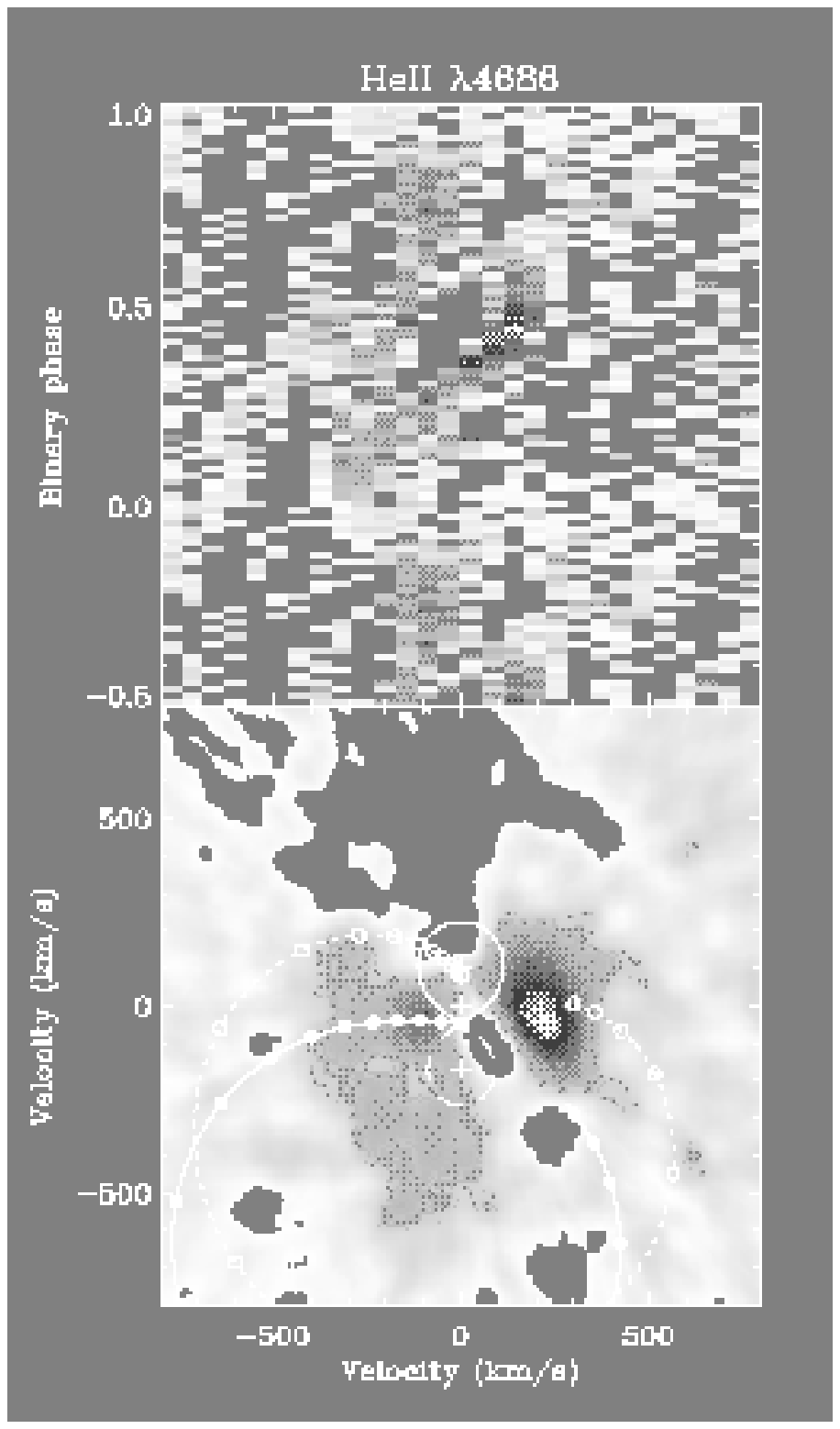}
\end{center}
\caption[]
{The He{\small~II} \index{s}{Doppler map}Doppler map of
\index{o}{Her X-1}Her X-1 showing  a spot of emission
close to    the  \index{s}{neutron  star}neutron  star   at velocities
associated to the   \index{s}{gas stream}gas stream.    The parameters
used  were  q=1.67   (mass of donor/accretor)     and M=1.3  for   the
\index{s}{neutron star}neutron star \cite{quaintrell00}.}
\label{HeII_map}
\end{figure}

\clearpage

\begin{figure}[h]
\begin{center}
\resizebox{\hsize}{!}{\rotatebox{-90}{\includegraphics[width=.8\textwidth]{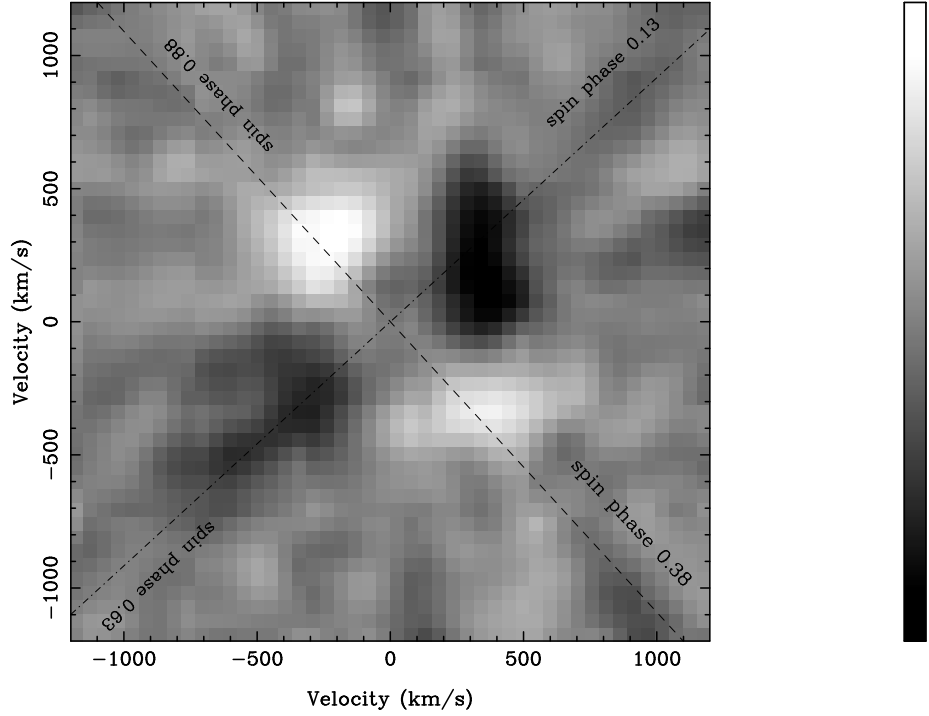}}}
\end{center}
\caption[]
{The     \index{s}{Doppler map}Doppler     map   of  the  double-pulse
He{\small~II} emission profile  with \index{s}{spin period}spin period
(545   seconds)  of the   intermediate  polar \index{o}{RX J0558+53}RX
J0558,  as viewed from the  compact object (0,0).   Back projection of
the He{\small~II}  pulsed emission profiles produces a quadrapole-like
velocity distribution, consisting of the minima ('dark' shade) and the
maxima ('bright' shade) of the two pulses.  The  spin phases where the
above  are more pronounced are  also marked.  The  emission line pulse
lags behind the continuum pulse by  0.12 spin cycles giving a powerful
insight  into the coupling region  between the Kepler-orbiting gas and
the magnetosphere \cite{harlaftis99a}.}
\label{j0558}
\end{figure}

Perhaps, the  magnetic \index{s}{magnetic propeller}propeller model is
the  likely interpretation  of  the  He{\small~II}   \index{s}{Doppler
map}Doppler   map of\index{o}{Her X-1}   Her X-1  where a low-velocity
emission spot  close  to the  \index{s}{neutron star}neutron  star  is
revealed   (Fig.~\ref{HeII_map}).   Alternatively,   the    1.24   sec
searchlight  beam from    the  \index{s}{neutron   star}neutron   star
illuminates the truncated inner disc.  Gas  from the inner edge of the
disc is then funneled along the magnetic field lines onto the poles of
the neutron star (for a review see
\cite{patterson94}  and references therein).  In   this case, there is
current  consensus that  the  disc  feeds  gas  to  the compact object
through an `accretion curtain' model \index{s}{accretion!curtain}
\cite{ferrario88,rosen88} which produces  a dipole pattern in both
emission and absorption as the searchlight beam  from each pole passes
through the  \index{s}{accretion!curtain}accretion curtain to the line
of    sight  \cite{harlaftis99a}.    For   example, the   double-pulse
He{\small~II} emission    profile coming from    the 545-sec  spinning
compact  object   in  the \index{s}{cataclysmic  variable!intermediate
polar}intermediate  polar \index{o}{RX J0558+53}RX  J0558+53 is mapped
as such  a dipole pattern in both  emission and absorption  centred on
the white dwarf (Fig.~\ref{j0558}).

\begin{figure}[bh]
\begin{center}
\resizebox{\hsize}{!}{\rotatebox{-90}{\includegraphics[width=.8\textwidth]{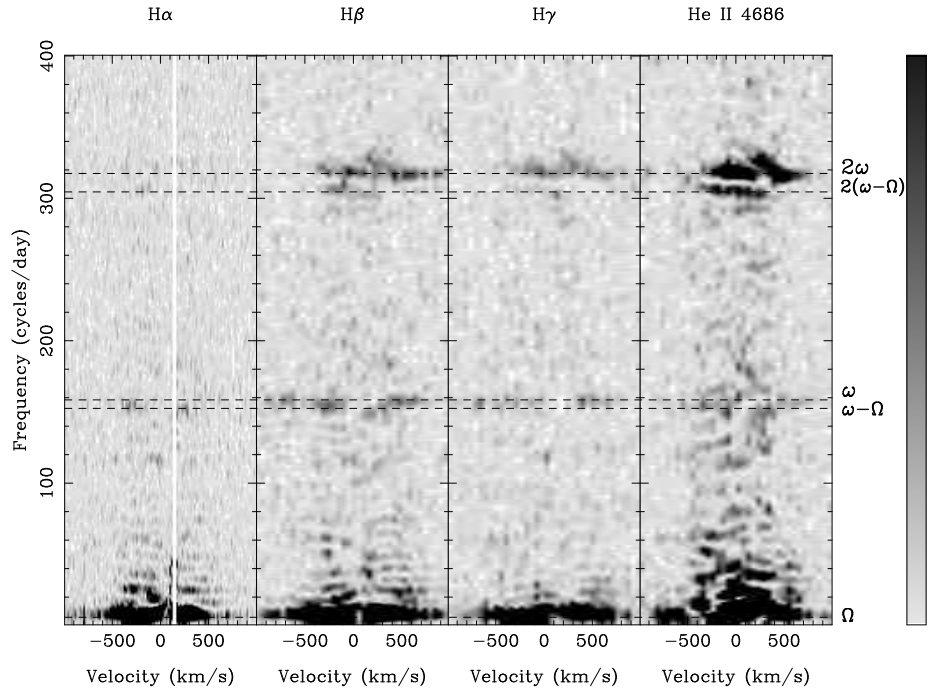}}}
\end{center}
\caption[]{The Fourier periodogram per velocity bin in the continuum 
and the emission lines. See text.The spin frequency is only evident in
H$\beta$ whereas the first  harmonic in dominant  in all power spectra
except that of H$\alpha$.  An orbital side-band at $2~(\omega-\Omega)$
is also clearly present\index{o}{RX J0558+53}.}
\label{j0558b}
\end{figure}

The  illuminating  effects of   the   compact object's  beams  on  the
surrounding supersonic gas  can provide insight in  the inner  disc of
X-ray binaries through periodogram analysis of the line profiles.  For
example,   harmonics of the  beat frequency  between the $\omega$ spin
frequency and the  $\Omega$  orbital frequency  as well  as  different
combinations   of these frequencies   are then  suggestive of specific
illuminating patterns.   For example, a  simple, disc-fed emission has
most power in the $2\omega$ frequency.  Such a periodogram analysis of
power spectra of line profiles against frequency and velocity is shown
in Fig.~\ref{j0558b} where prevalence  of the $2~(\omega-~\Omega)$ and
$2~\omega$ frequencies   indicate both disc-   and stream-fed emission
from two  diametrically-opposed poles with similar emission properties
a  truncated  disc  is implied  in  \index{o}{RX  J0558+53}RX J0558+53
\cite{harlaftis99a}.  This analysis  provides a powerful  probing tool
in the coupling  region between the
\index{s}{supersonic}supersonic gas   in    the inner disc   and   the
magnetosphere   of the   compact object   and   perhaps  this will  be
undertaken soon for the He{\small~II} line of Her X-1 \index{o}{Her X-1}.

\clearpage

\section{\bf Future prospects}

Indirect imaging techniques,  utilizing optical  spectroscopy, can now
probe \index{s}{X-ray  binary}X-ray   binary physics  with  sufficient
signal   to  noise ratio and  start   distinguishing accretion details
comparable to those observed in the brighter
\index{s}{cataclysmic variable}cataclysmic variables.
Moreover,  the   quality of  X-ray  observations   is  such  now  that
applications in  this domain is the next  step forward.  The behaviour
of the hard \index{s}{X-ray  emission}X-rays with respect to  the soft
\index{s}{X-ray emission}X-rays (time lags and  spectra) can probe the
size  of the  Compton  scattering  region   and infer radial   density
profiles
\cite{hua99}.  Image reconstruction  of   the hot electron  \index{s}{plasma}plasma 
may result in defining better properties of the hot accretion disc and
clarify its relation to advection dominated accretion flows.  The
spectral resolution  of iron profiles  has considerably increased with
the advent of the ASCA satellite and has  revealed in many interactive
binary  systems three  peaks  in   the  iron profile,  namely  thermal
emission at 6.7 and 7.0 KeV, and fluorescent emission at 6.4 KeV
\cite{ezuka99}.  \index{s}{Doppler tomography}Doppler tomography and
echo tomography using X-ray iron line profiles and continuum, and even
\index{s}{eclipse  map}eclipse maps of the  accretion  disc from X-ray
light curves \cite{mukai97}, may  become possible  with the new  X-ray
satellites.

%
\clearpage
\addcontentsline{toc}{section}{Objects Index}
\flushbottom
\printindex{o}{Objects index}
\addcontentsline{toc}{section}{Index}
\printindex{s}{Index}

\end{document}